\documentclass[12pt]{article}
\usepackage{amsmath,amssymb,fullpage}
\usepackage{natbib}

\begin{document}
\begin{center}
\textbf{\large{Discussion on Computationally Efficient Multivariate Spatio-Temporal Models for High-Dimensional Count-Valued Data by Jonathan R. Bradley, Scott H. Holan, and Christopher K. Wikle}}\\ \vspace{10pt}
\textbf{William Weimin Yoo}\\
\textit{Leiden University}
\end{center}
\begin{abstract}
I begin my discussion by summarizing the methodology proposed and new distributional results on multivariate log-Gamma derived in the paper. Then, I draw an interesting connection between their work with mean field variational Bayes. Lastly, I make some comments on the simulation results and the performance of the proposed Poisson multivariate spatio-temporal mixed effects model (P-MSTM).
\end{abstract}

\noindent\textbf{Keywords:} Multivariate log-Gamma, Spatio-temporal, Variational Bayes, Mean field, Latent Gaussian Process\vspace{10pt}\\

I would like to congratulate the authors for such an interesting and important work in spatio-temporal statistics. Indeed, high-dimensional count-valued data is a norm in large-scale census studies across the world, and the authors proposed an efficient and innovative procedure to model this complex data that scales well with its size. Let me briefly summarize their methodology before I begin my discussion. At the highest hierarchy, counts are modelled using a Poisson distribution and the log-link is used to link its mean with the underlying latent process. This latent process in turn has a mixed effects model representation, where the fixed effect is a linear combination of spatio-temporal covariates and the random effect part consisting of spatio-temporal basis functions. The authors took a departure from the Latent Gaussian Process approach (widely regarded as the industry standard), by modeling the fixed and random effects coefficients with multivariate log-Gamma (MLG) priors.

As its name suggests, the log-Gamma is simply the logarithm of a Gamma distributed random variable. The authors then took the opportunity to develop new distributional theory for MLG's. In particular, they derived probability density function of MLG under affine transformation and also their conditional distributions. The most striking result here is Theorem 2, where they established equivalence between the conditional MLG to certain classes of marginal MLG. This then enables them to sample efficiently from conditional MLG's and they designed a fast Gibbs sampler based on this new sampling scheme.

The strategy of reducing the simulation of a complicated conditional MLG to simulation using its equivalent marginal distribution, is reminiscence to another class of methods called Variational Bayes (VB) used especially in the machine learning community for massive data problems. As opposed to MCMC algorithms, VB seeks an analytic approximation to the posterior such that this approximation is close to the posterior in Kullback-Leibler divergence. A widely used strategy in VB is to assume that the approximating multivariate distribution has a factorised form (mean field VB), e.g., product of marginals across parameters and latent variables. It is conceivable that for a Poisson likelihood and MLG priors as considered in this paper, the resulting best approximating marginals for the parameters will also be a MLG due to conjugacy and I think they will have the same form as in (10) of Theorem 2 in \citet{discuss}. As a result, the mean field VB will also produce an iterative procedure much like the Gibbs sampler algorithm proposed by the authors, but it will be a set of circular equations updating the hyperparameters (scale and rate) of the MLG marginal approximations. Although there are not much theory on VB, but empirical studies in real-world massive data applications seem to show that VB has comparable estimation performance as MCMC and is several magnitudes faster (\citet{jordan}).

My other comment centers around the out-of-sample simulation experiment. In Figure 1, the proposed model captures the global spatial pattern well but seems to underestimate regions with high employment, and I was wondering how could one fine-tune the model to better capture these local county-level characteristics. A very natural idea is to include economic indicators for a county (if available) or some seasonality correction term in the fixed effect covariates, since employment numbers depend on economic/commercial activities of a county and they tend to follow job market seasons. My last point is about the performance between the proposed P-MSTM model and the ``industry standard" LGP (Latent Gaussian Process). It was discussed in the paper that LGP is inefficient compared to P-MSTM, but I am also curious about the predictive performance of P-MSTM in comparison to LGP for the simulation and the actual data analysis considered in this paper. In particular, whether P-MSTM achieves the same accuracy as LGP using much less computer running time.

Massive and high-dimensional data is now a norm in spatio-temporal statistics, and this paper, through the development of new distributional theory, opens up a way to model and compute these complex datasets. Interesting future work might include generalizing the proposed modeling strategy to encompass both count (discrete) and continuous data. I envision that this paper will inspire new research activities by encouraging statisticians to explore models beyond Gaussian Process and stationarity.

\bibliographystyle{apa}
\bibliography{spatial}

\end{document}